\documentclass[smus]{snow2e}
\usepackage{graphics}

\begin{document}
\setlength{\titleblockheight}{5cm}

\title{
\begin{flushright}
{\large{FERMILAB-Conf-96/431}}\\
{\large{November 1996}}\\
\end{flushright}
Discovering Technicolor at Hadron Colliders\thanks
{To appear in the Proceedings of the 1996 DPF/DPB Summer Study on
New Directions for High Energy Physics (Snowmass 96).}}

\author{John~Womersley\\ {\textit{Fermi National Accelerator
Laboratory, Batavia, IL 60510}}}

\maketitle

\thispagestyle{empty}\pagestyle{empty}

\begin{abstract} 
Strategies are presented for discovering light,
color-singlet technipions ($\pi_T$), produced in association with a vector
boson through $s$-channel
technirho production, at the Tevatron and LHC.
Signal and $W+$jets background were simulated including detector
effects. Tagging of $b$-quarks from the $\pi_T \to b\overline b$ decay
is found to be important to reduce the $W+$jets background.    
The kinematic properties of signal and background 
events are significantly different and simple cuts can be used to further
improve the signal to background ratio.  
\end{abstract}

\section{Introduction}
Discovery strategies for light, color singlet technipions
at hadron colliders have been investigated.
As pointed out by Eichten and Lane\cite{eichtenlane}
they can be copiously produced through $s$-channel 
technirho production:
$$ q \overline q^\prime \to \rho^\pm_{T} \to V_1 V_2 $$
where $V_1 V_2 = W^\pm Z$, $W^\pm \pi^0_T$, $\pi^\pm_T Z$,
or $\pi^\pm_T \pi^0_T$; and through
$$  q \overline q \to \rho^0_{T} \to V_1 V_2 $$
where $V_1 V_2 = W^+ W^-$, $W^\pm \pi^\mp_T$,
or $\pi^+_T \pi^-_T$.

The modes where $W$ and $Z$ are produced and subsequently
detected in their leptonic decays are straightforward and
largely free of background; see for example the ATLAS 
Technical proposal\cite{atlastp}.
In this study, the dijet decays of the technipion have
been investigated:
\begin{eqnarray*}
\pi^0_T & \to & b\overline b \\
\pi^\pm_T & \to & c \overline b
\end{eqnarray*}
These will generally be expected to dominate as
long as the $t \overline t$ and $ t \overline b$
modes are kinematically accessible; in Topcolor
models the top decay modes can remain forbidden even for
larger masses.    

\section{Signal and Background}

For definiteness the following process has been considered:
$$ q \overline q^\prime \to \rho_T \to W(\ell \nu) \pi_T (b\overline b),$$
with $m_{\rho_T} = 210\,$GeV and $m_{\pi_T}=115\,$GeV. 
The signal is thus a $W$ 
(reconstructed from lepton plus missing transverse energy)
together with two jets, with a resonance in the dijet mass $m_{jj}$.
The backgrounds are $W+$jets and $t\overline t$.  (The latter has not yet been
included in the study since it is small compared with the $W+$jets 
process for $n=2$ jets, even if single $b$-tagging is applied).  
The signal cross sections are large: about 5~pb at the Tevatron, and
35~pb at the LHC\cite{lane}.  
The main issue is therefore dijet mass resolution and
$b$-tagging.  

Signal and background events were simulated using ISAJET.  The signal
topology was generated using the TCOLOR process with the $WZ$ final state;
the $Z$ mass was set to $m_{\pi_T}$ and the decay to $b\overline b$ was
forced.

\begin{figure}[t]
\leavevmode
\begin{center}
\vspace{2.5cm}
\resizebox{!}{8cm}{%
\includegraphics{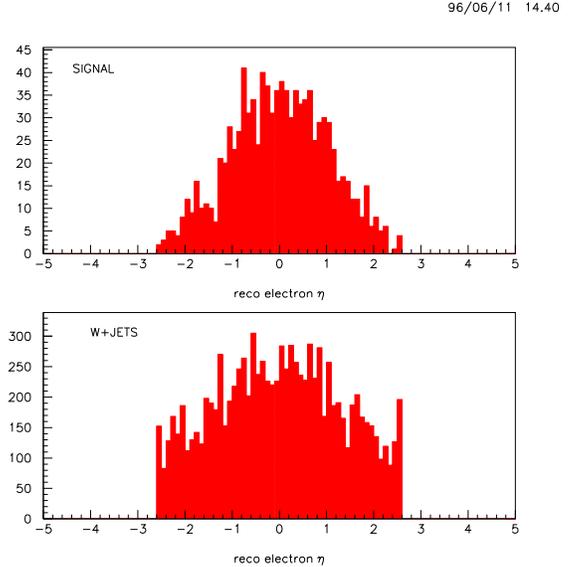}}
\end{center}
\vspace{-2.5cm}
\caption{Reconstructed lepton pseudorapidity distributions for signal
and $W+$jets processes.}
\label{fig:eta_e}
\end{figure}

Detector acceptance and resolution were modelled using a fast 
simulation\cite{sscsim}.
Energy was deposited in cells of size $\Delta\eta \times \Delta\phi =
0.1 \times 0.1$ and was smeared for
detector resolution: $15\%/\sqrt{E{\rm(GeV)}} \oplus 0.5\%$ for EM, and
$50\%/\sqrt{E{\rm(GeV)}} \oplus 5\%$ for hadronic energy.
Transverse shower spreading and calorimeter leakage
were also modeled.  Jets were found (up to $|\eta| = 4$)
from the calorimeter towers using a cone of $R=0.7$. Missing transverse
energy was calculated from the sum of the calorimeter towers over 
$|\eta| \leq 5$.

Events were selected which satisfied the following criteria:
\begin{itemize}
\item A good $W \to \ell \nu$ candidate, defined as:
\begin{itemize}
\item[$\circ$] lepton with $p_T^\ell > 25$~GeV/c, $|\eta^\ell| < 1.1$, and
	isolated (transverse energy within $R < 0.4$ less than
	10\% of the lepton $p_T$);
\item[$\circ$] $E_T^{miss} > 25$GeV
\item[$\circ$] Transverse mass $m_T$ satisfying $50 < m_T < 100$~GeV;
\end{itemize}
\item At least two jets with $E_T > 20$~GeV and $|\eta^j|<2.5$.
\end{itemize}
The lepton was required to be central, since (as Fig.~\ref{fig:eta_e} shows)
this gives some improvement in signal-to-background.
For $b$-tagging it was assumed that single tagging would be performed
with an efficiency of 50\% and a mistag rate of 1\% for light quark jets.

The cross sections obtained for the Tevatron and LHC are listed in
Table~\ref{table}.
It will be seen that the signal-to-background ratio 
for this process is rather better at the Tevatron. 

\begin{figure}[t]
\leavevmode
\begin{center}
\vspace{2.5cm}
\resizebox{!}{8cm}{%
\includegraphics{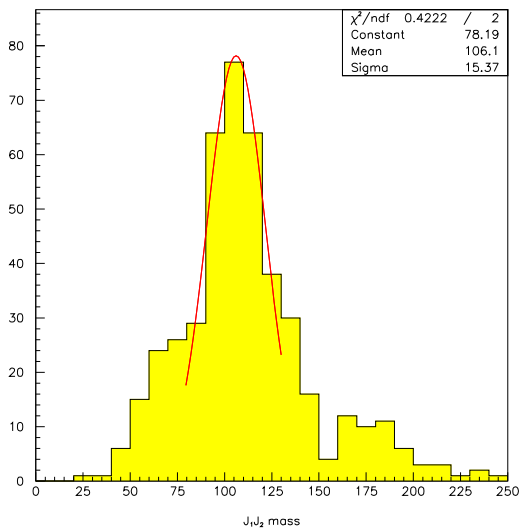}}
\end{center}
\vspace{-2.5cm}
\caption{Leading dijet invariant mass distribution for
$W(\ell\nu)\pi_T(b \overline b)$ events at the LHC.}
\label{fig:jjmass}
\end{figure}

Figure~\ref{fig:jjmass} shows the invariant mass distribution obtained 
for the leading jet pair in signal events. The peak has a
resolution of about 15~GeV with tails from jet combinatorics.     

\begin{figure}[t]
\leavevmode
\begin{center}
\vspace{2.5cm}
\resizebox{!}{8cm}{%
\includegraphics{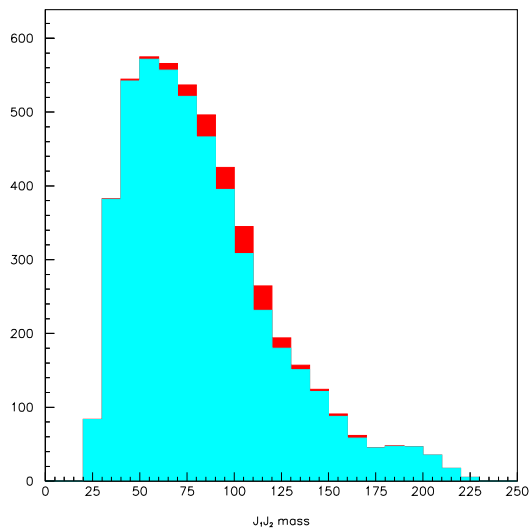}}
\end{center}
\vspace{-2.5cm}
\caption{Leading dijet invariant mass distribution for
technipion signal (dark) over the  $W+$jets background 
(light) at the Tevatron, before $b$-tagging. Vertical
scale is events/10~GeV/2~fb$^{-1}$.  The background has
been smoothed to simulate the full statistics.}
\label{fig:tev1}
\end{figure}

\begin{figure}[t]
\leavevmode
\begin{center}
\vspace{2.5cm}
\resizebox{!}{8cm}{%
\includegraphics{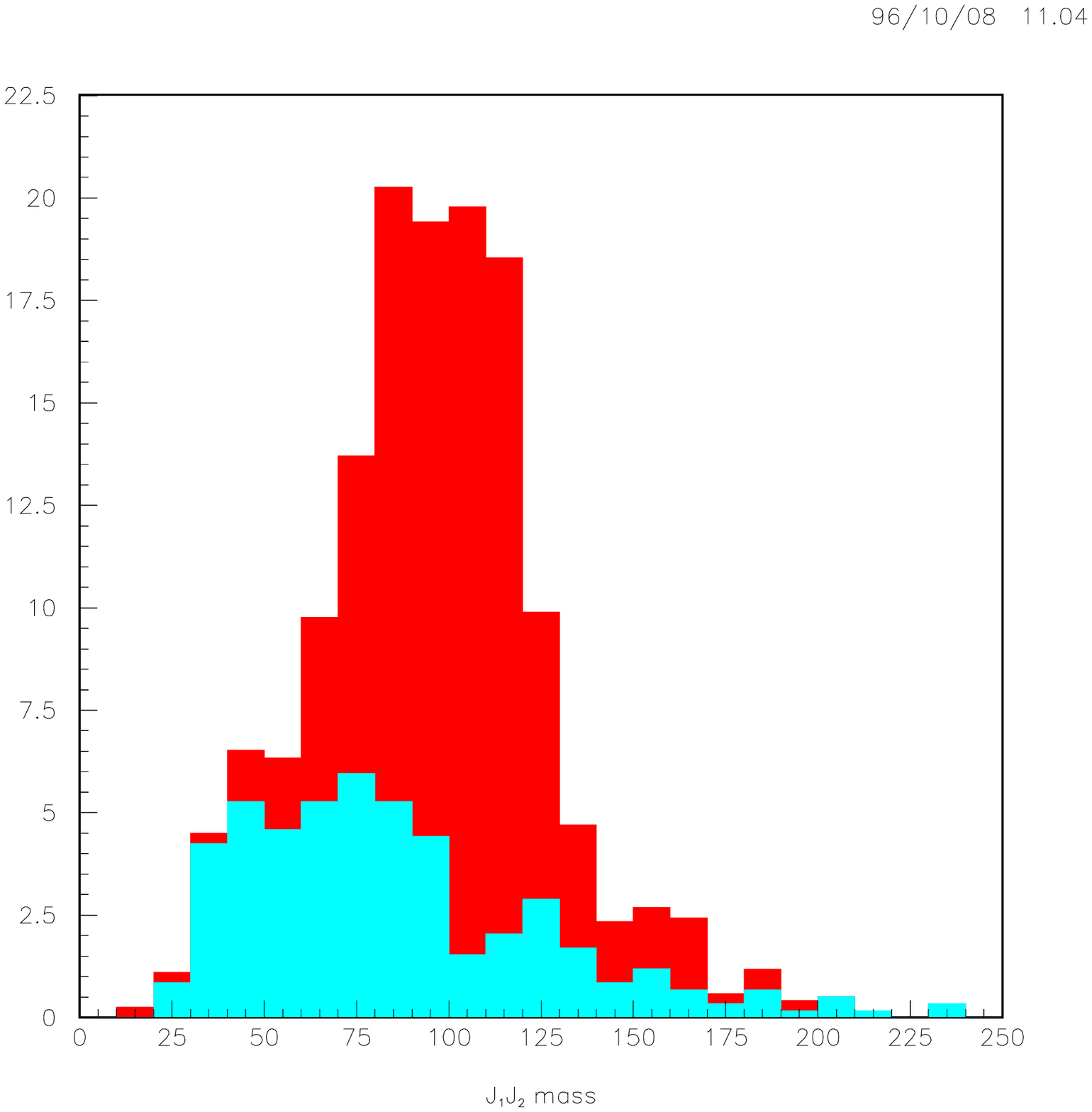}}
\end{center}
\vspace{-2.5cm}
\caption{Leading dijet invariant mass distribution for
technipion signal (dark) over the  $W+$jets background 
(light) at the Tevatron, after $b$-tagging. Vertical
scale is events/10~GeV/2~fb$^{-1}$.}
\label{fig:tev2}
\end{figure}

\begin{figure}[t]
\leavevmode
\begin{center}
\vspace{2.5cm}
\resizebox{!}{8cm}{%
\includegraphics{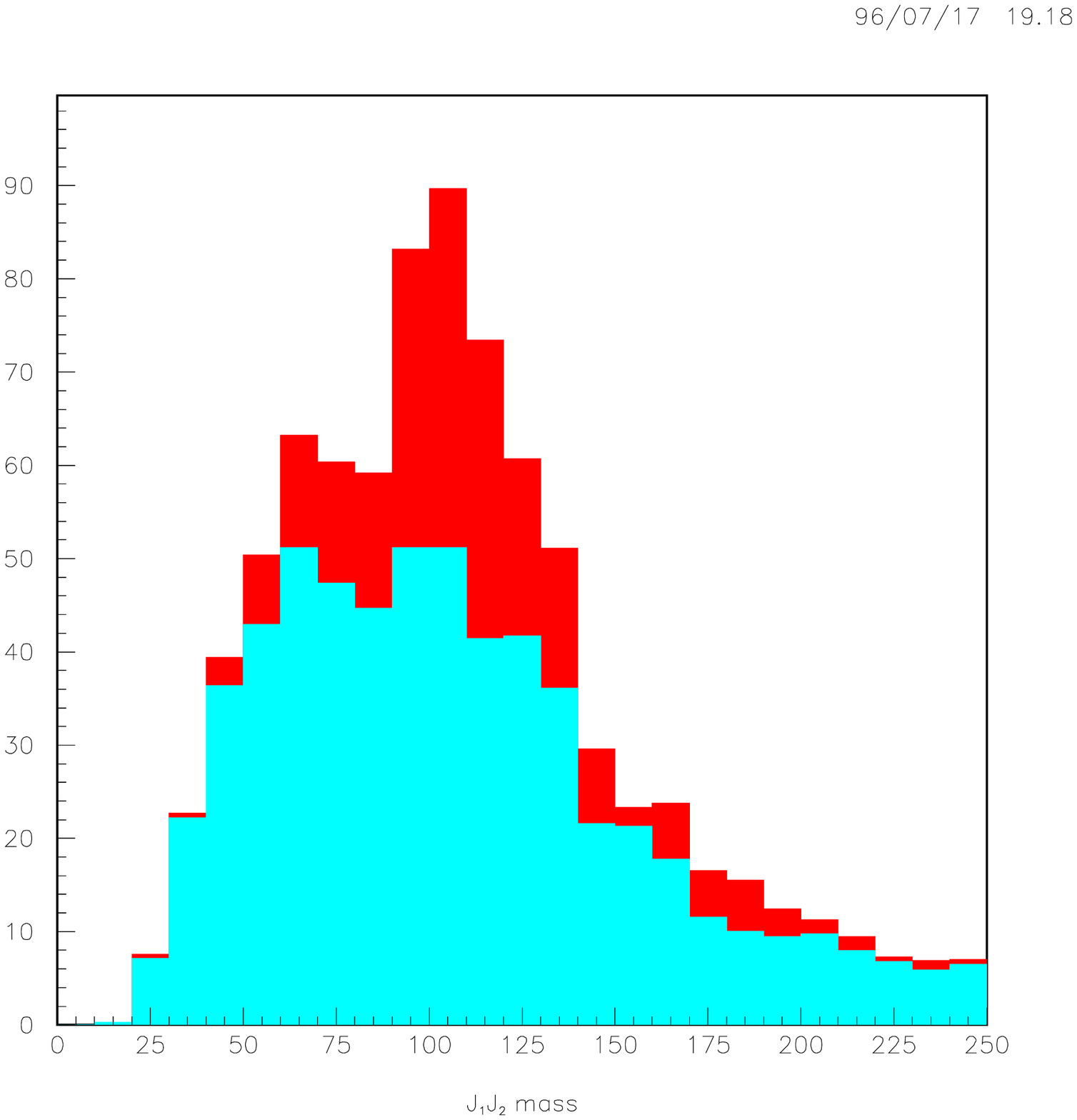}}
\end{center}
\vspace{-2.5cm}
\caption{Leading dijet invariant mass distribution for
technipion signal (dark) over the  $W+$jets background 
(light) at the LHC, after $b$-tagging. Vertical
scale is events/10~GeV/0.5~fb$^{-1}$.}
\label{fig:lhc2}
\end{figure}
\begin{table}[h]
\begin{center}
\caption{Cross sections for signal and $W+$jets background.}
\label{table}
\begin{tabular}{lcc}
\hline
\hline
  &$W \pi_T$ &$W+$jets \\
\hline
LHC \\
\hline
$\sigma\cdot B$ &3.7~pb  &2200~pb\\
with 2 jets     &1.7~pb  &250~pb\\
with $b$-tag    &0.85~pb &2.5~pb\\
\hline
Tevatron \\
\hline
$\sigma\cdot B$ &0.53~pb  &170~pb\\
with 2 jets     &0.10~pb  &2.5~pb\\
with $b$-tag    &0.05~pb  &0.025~pb\\
\hline
\hline
\end{tabular}
\end{center}
\end{table}

Figure~\ref{fig:tev1} shows the invariant mass distribution obtained 
for signal and background at the Tevatron, before $b$-tagging.
Once $b$-tagging is applied, the signal becomes much more apparent,
as seen in Fig.~\ref{fig:tev2}.  A clear excess is visible with $S:B\sim 5$
in the peak, and could easily be discovered in Run II with 2~fb$^-1$.
For comparison, Fig.~\ref{fig:lhc2} shows the situation at the LHC after
$b$-tagging.

\section{Kinematic Properties of the Events}

We note that there are significant differences in kinematic 
distributions between signal and background events.
Cuts on these distributions may be used to further improve the 
background rejection,
or they may be used as a way to confirm the 
presence of a signal by (for example) observing
differences in these variables as a function of
dijet mass.

\begin{figure}[t]
\leavevmode
\begin{center}
\vspace{2.5cm}
\resizebox{!}{8cm}{%
\includegraphics{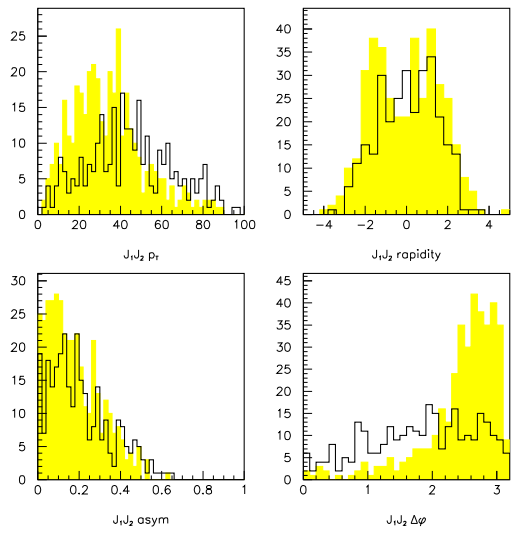}}
\end{center}
\vspace{-2.5cm}
\caption{Distributions of $p_T^{jj}$, $\eta^{jj}$, asymmetry
$A$ and $\Delta\phi$ for
technipion signal (shaded) and  $W+$jets background 
(outline) at the Tevatron.}
\label{fig:kinematics}
\vspace{2.85cm}
\end{figure}

Some variables of interest are:
\begin{itemize}
\item
Transverse momentum of the leading dijet system, $p_T^{jj}$;
\item
Pseudorapidity of the leading dijet system, $\eta^{jj}$;
\item
$\Delta\phi$ between the leading two jets;
\item
Dijet asymmetry $A = (E_{T1}-E_{T2})/(E_{T1} + E_{T2})$.
\end{itemize}
The dijet pseudorapidity and asymmetry do not offer much discriminating
potential, but the $\Delta\phi$ and transverse momentum of the dijet system
are distinctly different between signal and background,
as can be seen from Fig.\ref{fig:kinematics}.
Requiring, for example, $\Delta\phi > 2.3$, and $p_T^{jj} < 45$~GeV/c 
retains 60\% of the signal while rejecting 74\% of 
the $W+$jets background.

\section{Conclusions}

Light, color-singlet technipions, produced in association with a vector
boson through s-channel
technirho production, can be discovered at hadron colliders
in the $b\overline b$ decay mode. The signal to background ratio
is somehat better at the Tevatron but this physics can also be addressed
at the LHC.  Tagging of $b$-quarks is important to reduce the $W+$jets
background.    
The kinematic properties of signal and background 
events are significantly different and simple cuts can be used to further
improve the signal to background ratio.

%

\vfill

\end{document}